\documentstyle[NATO,numreferences,epsf]{CRCKAPB}

\newcommand{\msun}{{M}_{\odot}}

\newcommand{\neon}{^{22}\mathrm{Ne}\;}
\newcommand{\Ne}{^{22}\mathrm{Ne}\;}

\newcommand{\carb}{^{12}\mathrm{C}\;}
\newcommand{\oxy}{^{16}\mathrm{O}\;}
\newcommand{\oxyet}{^{18}\mathrm{O}\;}
\newcommand{\nitr}{^{14}\mathrm{N}\;}
\newcommand{\flor}{^{18}\mathrm{F}\;}
\newcommand{\xne}{X_{22}}

\renewcommand{\vec}[1]{{\mathbf #1}}
\newcommand{\rhat}{\vec{\hat{r}}}

\newcommand{\der}[2]{\frac{d \,#1}{d#2}}

\newcommand{\pder}[2]{\frac{\partial \,#1}{\partial#2}}

\begin{opening}
\title{SINKING OF $^{22}$Ne IN LIQUID WD INTERIORS}

\author{C. J. DELOYE}
\institute{Department of Physics, University of California, Santa Barbara Broida Hall, Santa Barbara, CA 93106}
\author{L. BILDSTEN}
\institute{Kavli Institute for Theoretical Physics and Department of Physics, University of California, Santa Barbara\\Kohn Hall, Santa Barbara, CA 93106}
\end{opening}

\begin{document}
\begin{abstract}
We assess the impact of the trace element $\neon$ on the cooling and seismology of a liquid C/O white dwarf (WD).  Due to this elements' neutron excess, it sinks towards the interior as the liquid WD cools. The gravitational energy released slows the WD's cooling by 0.5-1.6 Gyr. In addition, the $\neon$ abundance gradient changes the periods of the high radial order $g$-modes at the 1\% level.  
\end{abstract}

\section{Biasing of $\neon$ Diffusion in Liquid White Dwarf Interiors}
After $\carb$ and $\oxy$, the most abundant nucleus in a $M<\msun$ white dwarf (WD) interior is $\neon$.  As the slowest step in the CNO cycle is the $p+\nitr$ reaction, almost all CNO nuclei end up as $\nitr$ at the end of H burning.  During the He burning stage, the reaction $\nitr(\alpha\,,\gamma)\flor(\beta^+)\oxyet(\alpha,\gamma)\neon$ processes all $\nitr$ into $\neon$.  For recently formed WDs of $M<\msun$ this results in a $\neon$ mass fraction $\xne \approx Z_{\mathrm{CNO}}\approx 0.02$ (see, for example, \cite{um99}).

$\neon$ has a mass to charge ratio, $A/Z$ greater than that of $\carb$ and $\oxy$.    As pointed out by \cite{brav92} and \cite{bh01}, this affects the dynamics of $\neon$ nuclei in the following manner.  In the degenerate WD interior, there exists an upward pointing electric field of magnitude $eE\approx 2m_p g$ \cite{bh01}, $g$ being the local gravitational acceleration.  The resulting net force on $\neon$, $\vec{F}=- 2m_pg \rhat$, biases $\neon$'s diffusion inward.  In contrast, the predominant $\carb$ and $\oxy$ ions experience no net force.  The resulting sedimentation of $\neon$ will impact the cooling history and seismology of the WD.

The $\neon$ evolution is governed by mass continuity
\begin{equation} \label{eq:cont}
\pder{\rho_{22}}{t} + \nabla\cdot {\bf J}_{22} =0 \;,
\end{equation}
where $\rho_{22}$ and ${\bf J}_{22}$ are the local $\neon$ density and flux respectively.  The flux is given by
\begin{equation} \label{:flux}
{\bf J}_{22} = (- D \pder{\rho_{22}}{r} - v \rho_{22}) \rhat\;,
\end{equation}
where  $D$ is $^{22}\mathrm{Ne}$'s diffusion coefficient in the background plasma and $v$ is the magnitude of its local drift velocity. $D$ and $v$ are related via the Stokes-Einstein equation: $v = 2 m_p g D/k T$.
 
As discussed in \cite{bh01} and \cite{del02}, $D$ is not well known for conditions in the WD interior.  So, for this work, we took as a starting point the self diffusion coefficient of the one-component plasma, $D_s$, as calculated in \cite{hans75},$D_s\approx 3 \omega_p a^2 \Gamma^{-4/3}$, where $\omega_p^2=4\pi n_i (Ze)^2/Am_p$, $\Gamma = (Z e)^2/a k T$ and $a^3=3A m_p/4\pi \rho$ is the inter-ion spacing of the background ions. At $\Gamma = 173$, the plasma crystallizes \cite{far93}.  We quantify the effects of the uncertainty of $D$ on our results by performing calculations with varying $D$.

\section{$\neon$ Sedimentation and WD Cooling}
We simultaneously evolve the initially constant $\xne$ and WD core temperature, $T_c$.  The $\neon$ evolution is modeled as above. We halt sedimentation in the solid phase.  We consider the WD interior to be isothermal, and calculate $T_c(t)$ using 
\begin{equation}
\label{eq:Tc}
\der{T_c}{t}\,\int_0^M \frac{C_V}{A m_p} dm = -L + L_g + \frac{l}{A m_p} \der{M_{crys}}{t}\;,
\end{equation}
where $C_V$ is the heat capacity, $L$ is the WD luminosity, $L_g$ the power generated by $\neon$ sinking, $l$ is the latent heat released per nucleon, and $M_{crys}$ is the mass of the WD that has crystallized. $L$ is determined from the $L-T_c$ relations of \cite{ab98} and $L_g = \int_0^R F (J_{22}/\rho_{22}) n_{22} 4 \pi r^2 dr$, where $n_{22} = \rho_{22}/22 m_p$. We ignore fractionation effects.  Finally, throughout we utilize background WD models composed of a single ion species, representing realistic C/O WD compositions by taking $A=14$ and $Z=7$.

\begin{figure}
\vspace{-.1cm}
\epsfxsize=3.0in
\epsfbox{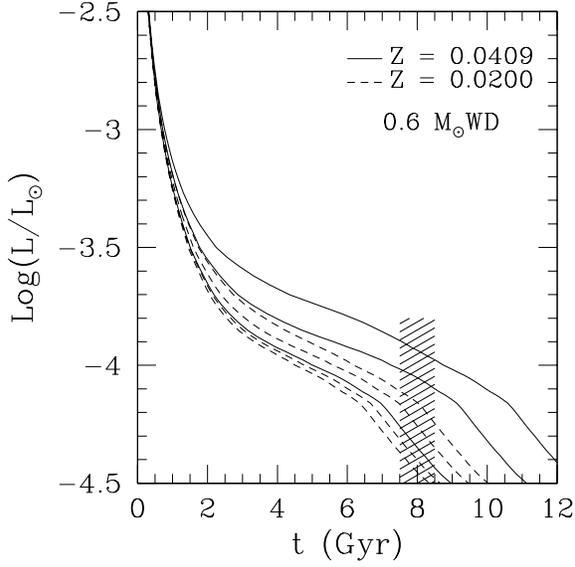}
\caption{Cooling of a $0.6 \msun$ WD.  The lower dashed curve shows the results neglecting $\neon$. The remaining curves show (from bottom to top) the results for $D=0,\,D_s,\,5 D_s,$ and $10 D_s$ for the specified $\xne$ value. The shaded regions gives the age range of NGC 6791. \label{fig:LtZcomp}}
\vspace{-.1cm}
\end{figure}

The overall impact of $\neon$ sedimentation on the WD age depends on the diffusion rate and when the WD crystallizes.  Sinking is fast in massive WDs, but crystallization occurs earlier, somewhat mitigating the impact.  The opposite is the case for low mass WDs, where sedimentation is slower, but can continue for up to 10 Gyrs.  $\neon$ content is also a factor, as $L_g$ depends on the $\xne$.  
\begin{figure}
\vspace{-.4cm}
\epsfxsize=3.0in
\epsfbox{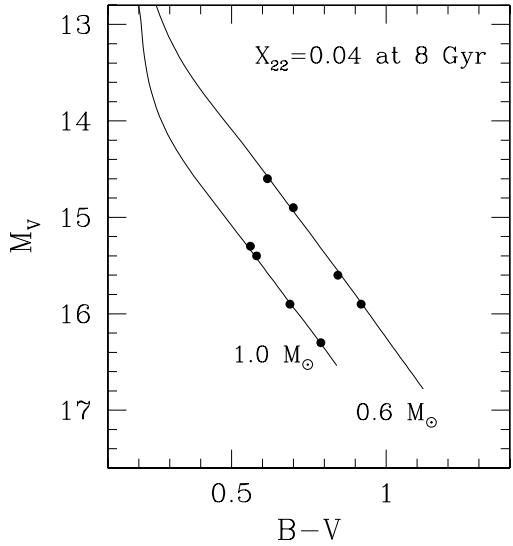}
\caption{Theoretical CMD for NGC 6791, showing the cooling tracks for $0.6$ and $1.0 \msun$ WDs.  Locations of WDs today for (bottom to top) $D=0,\,D_s,\,5 D_s,\, 10 D_s$ are shown by the dots. \label{fig:obscon}}
\end{figure}
Figure \ref{fig:LtZcomp} shows cooling curves for a $0.6 \msun$ WD model, comparing differing values of $D$ and $\xne$.  The higher metallicity is that of NGC 6791, a metal rich, old ($8.0 \pm 0.5$ Gyr) open cluster \cite{chab99}.  Figure \ref{fig:obscon} details the expected CMD location of WDs in NGC 6791 today assuming several values of $D$, demonstrating how observations can constrain on the value of $D$.

\section{Enhanced Internal Buoyancy from $\neon$}
The gradient in $\neon$ abundance creates a gradient in the mean molecular weight per electron, $\mu_e$, that provides an extra restoring force for the g-modes in the WD interior.  We can approximate the Brunt-V\"{a}is\"{a}l\"{a} frequency, $N^2$, in this case by considering an isothermal plasma of degenerate, relativistic electrons  and ideal ion gas with a $\mu_e$ gradient due to $\xne$ gradients. Doing so, we find (see \cite{del02} for details)
\begin{equation} \label{eq:n2rel}
N^2 = \frac{3}{8}\frac{k T}{h^2 \mu_i m_p} - g \frac{1}{11} \der{X_{22}}{z} \;,
\end{equation}
where $h=P/\rho g$ and $\mu_i$ is the ion mean molecular weight.  The first piece is the thermal contribution to $N^2$  while the second is the $\neon$ contribution.  

We estimate the impact of $\neon$ on the mode frequencies by performing the integral $\int N h/r d(\ln P)$ through the liquid WD
interior (the envelope's contribution to the integral is $\approx 0.3 \ {\rm rad \ s^{-1}}$).
Neglecting the $\neon$ contribution to $N^2$ gives $0.073 \ {\rm rad s^{-1}}$ through the WD core, while taking into account the $\Ne$ gradient gives $0.079 (0.095) \ {\rm rad \ s^{-1}}$ for
$D=D_s (5 D_s)$ . Including the envelope contribution shows $\neon$ can impact mode periods by more than 1\%, a level over 100 times larger than the measurement errors (see, e.g., \cite{brad01}). Clearly, if meaningful results are to be inferred from WD pulsation studies, $\neon$ gradients need to be taken into account.

This work was supported by the NSF under Grants PHY99-07949, AST01-96422, and AST02-05956.  L. B. is a Cottrell Scholar of the Research Corporation.

\end{document}